\documentclass[fleqn,12pt,twoside]{article}
\usepackage{espcrc1}

\usepackage{graphicx}
\usepackage[figuresright]{rotating}

\newcommand{\AmS}{{\protect\the\textfont2
  A\kern-.1667em\lower.5ex\hbox{M}\kern-.125emS}}
\newcommand{\al}{\alpha}
\newcommand{\be}{\beta}\newcommand{\ga}{\gamma}\newcommand{\vrr}{\varrho}
\newcommand{\de}{\delta}

\newcommand{\ve}{\varepsilon}

\newcommand{\De}{\Delta}

\newcommand{\equ}{\equiv}
\newcommand{\mb}{\mbox}
\newcommand{\bs}{\begin{sloppypar}} \newcommand{\es}{\end{sloppypar}}
\newcommand{\p}[1]{(\ref{#1})}
\newcommand{\lab}{\label} 
\newcommand\beq{\begin{eqnarray}} \newcommand\eeq{\end{eqnarray}}
\newcommand\beqstar{\begin{eqnarray*}} \newcommand\eeqstar{\end{eqnarray*}}
\newcommand{\beqe}{\begin{equation}} \newcommand{\eeqe}{\end{equation}}
\newcommand{\bal}{\begin{align}}

\title{Pairing effects in low density domain of nuclear matter}

\author{A.~A.~Isayev\address[MCSD]{
Kharkov Institute of Physics and Technology,  61108, Kharkov,
Ukraine}%
        \thanks{A.I. is grateful for  support of Topical Program of APCTP during
his stay at Seoul.},
        S. I. Bastrukov\address[Ewha]
        {Dept. of Physics and Center for Space Science and
        Technology,\\
        Ewha Womans University, Seoul 120-750, Korea}\address[Kgp]{Center for
High Energy  Physics, Kyungpook National University, Daegu
702-701, Korea}\address{Joint Institute for Nuclear Research,
141980, Dubna, Russia},
        J. Yang\addressmark[Ewha]\addressmark[Kgp]\thanks{J.Y. was partially supported
        by Korea Research Foundation Grant (KRF-2001-041-D00052).}}

\begin{document}

\maketitle

\begin{abstract}
Using equations, governing $np$ pairing correlations in $S=1, T=0$
pairing channel (PRC 63 (2001) 021304(R)), it is shown that at low
densities equations for the energy gap in the spectrum of
quasiparticles and chemical potentials of protons and neutrons
allow solutions with negative chemical potential. This corresponds
to appearance of Bose--Einstein condensate (BEC) of deuterons in
low density region of nuclear matter.
\end{abstract}

\vspace{6mm}

 The
transition from BCS superconductivity to Bose--Einstein
condensation  occurs in a Fermi system, if either density is
decreased or the attractive interaction between fermions is
increased sufficiently.  Recently it was realized that this phase
transition takes place in symmetric nuclear matter, when $np$
Cooper pairs at higher densities go over to BEC of deuterons at
lower densities~\cite{AFRS,BLS}. During the phase transition the
chemical potential changes its sign at certain critical density
(Mott transition), approaching  half of the deuteron binding
energy at ultra low densities. Here for studying corresponding
phase transition in  asymmetric nuclear matter we shall use
equations, obtained in Ref.~\cite{AIPY} for description of $np$
pairing correlations in $S=1, T=0$ pairing channel:
  \beq \De({\bf k})
&=&-\frac{1}{V}\sum_{{\bf k}'}V({\bf k},{\bf k}')\frac{\De({\bf
k}')}{2E_{k'}}(1-f(E_{k'}^+)-f(E_{k'}^-)),\; \lab{8}
\\
\vrr&=&\frac{2}{V}\sum_{\bf
k}\Bigl(1-\frac{\ve_k}{E_k}[1-f(E_k^+)-f(E_k^-)]\Bigr)\equ\frac{2}{V}\sum_{\bf
k}n_k,\lab{10}\\
\al\vrr&=&\frac{2}{V}\sum_{\bf k}\Bigl(f(E_k^+)-f(E_k^-)\Bigr),\;
E_k^\pm=E_k\pm\mu_{03}=\sqrt{\varepsilon^2_k+\Delta^2({\bf
k})}\pm\mu_{03}, \lab{11}\eeq   where $f(E)$ is Fermi
distribution, $\varepsilon_k$ is the single--particle spectrum,
$\mu_{00}, \mu_{03}$ being half of a sum and half of a difference
of proton and neutron chemical potentials. Eq.~\p{8} is equation
for the energy gap $\Delta$ and Eqs.~\p{10}, \p{11} are equations
for the total density $\vrr=\vrr_p+\vrr_n$ and neutron excess
$\de\vrr=\vrr_n-\vrr_p\equ\al\vrr$ ($\al$ being the asymmetry
parameter). Then, introducing the anomalous density
$$\psi({\bf k})=<a^+_{p,k}a^+_{n,-k}>=\frac{\De({\bf
k})}{2E_k}\Bigl(1-f(E_k^+)-f(E_k^-)\Bigr)$$ and using Eq.~\p{10},
one can represent Eq.~\p{8} for the energy gap in the form
\beqe\frac{k^2}{m}\psi({\bf k})+(1-n_k)\sum_{{\bf k}'}V({\bf
k},{\bf k}')\psi({\bf k}')=2\mu_{00}\psi({\bf k}).\lab{12}\eeqe In
the limit of vanishing density, $n_k\rightarrow0$, Eq.~\p{12} goes
over into the Schr\"odinger equation for the deuteron bound
state~\cite{BLS,LNS}. Corresponding energy eigenvalue is equal to
$2\mu_{00}$.

 Further for numerical calculations we shall use the
effective zero range force, developed in Ref.~\cite{GSM} to
reproduce the pairing gap in $S=1,T=0$ pairing channel with Paris
NN potential: \beqe V({\bf r}_1,{\bf
r}_2)=v_0\Bigl\{1-\eta\biggl(\frac{ \vrr(\frac{{\bf r}_1+{\bf
r}_2}{2})}{\vrr_0}\biggr)^\ga\Bigr\}\de({\bf r}_1-{\bf
r}_2)\lab{9},\eeqe where $\vrr_0$ is the nuclear saturation
density ($\vrr_0=0.16\,\mb{fm}^{-3}$), $v_0,\eta,\be$ are some
adjustable parameters. Besides, in the gap equation~\p{8},
Eq.~\p{9} must be supplemented with a cut--off parameter $\ve_c$.
We utilize the following set of parameters:
$\eta=0,\,v_0=-530\,\mb{MeV}\cdot
\mb{fm}^3,\,m=m_G,\,\ve_c=60\,\mb{MeV}$, where $m_G$ is the
effective mass, corresponding to the Gogny force
\begin{figure}[thb]
\begin{flushleft}
\includegraphics[height=7.2cm,width=14.4cm,trim=26mm 152mm 39mm
69mm, draft=false,clip]{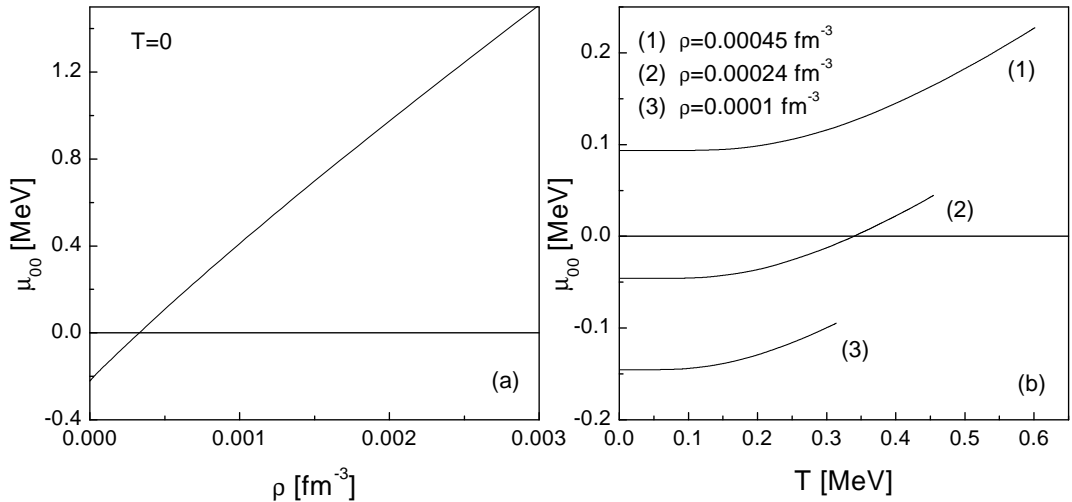}
\end{flushleft}
\caption{Chemical potential $\mu_{00}$ as a function of:
(a)~density at zero temperature, (b)~temperature at fixed
densities (low density region).}\label{fig1}
\end{figure}
D1S~\cite{BGG}. This set was developed to produce the bound state
between two extreme values: at zero energy and at the deuteron
binding energy~\cite{GSM}.

First we consider the case of symmetric nuclear matter ($\al=0$).
In Fig.~1(a) it is shown the zero temperature behavior of the
chemical potential $\mu_{00}$. It is seen that the chemical
potential at some density $\vrr_b$
($\vrr_b\approx3\cdot10^{-4}\,\mb{fm}^{-3}$) changes its sign,
that, according to Eq.~\p{12}, corresponds to appearance of
deuteron--like bound states in nuclear matter. When density tends
to zero, the chemical potential approaches its asymptotic value
$\mu_{00}=-\ve_b/2$,  $\ve_b$ being the binding energy of a bound
state. Therefore, we can conclude that under lowering density $np$
superfluidity smoothly evolves into the BEC of $np$ bound states
(deuterons).

In Fig.~1(b) we present the results of numerical determination of
the temperature dependence of the chemical potential $\mu_{00}$
for the fixed values of density. If density is low enough then
under decrease of temperature the chemical potential becomes
negative (the  curve 2), that
 corresponds to formation of bound states.
At very low densities $np$ condensate exists only in the form of
BEC of deuteron--like bound states (the curve~3). If densities is,
however, high enough, $np$ Cooper pairs survive even at zero
temperature (the curve 1).

\begin{figure}[tbp]
\begin{flushleft}
\includegraphics[height=7.0cm,width=8.6cm,trim=48mm 142mm
57mm 69mm, draft=false,clip]{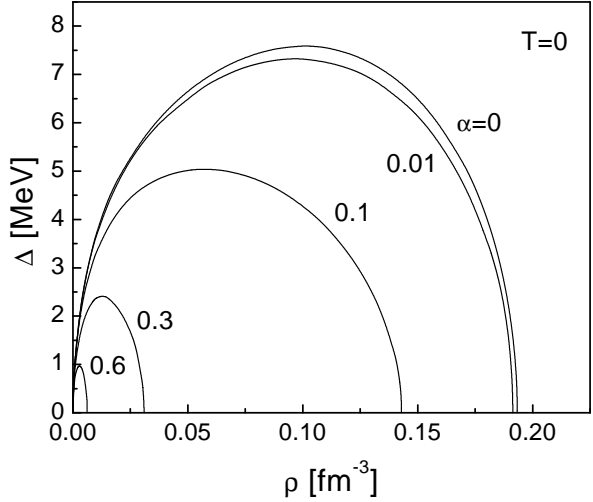}
\end{flushleft}
\caption{Energy gap as a function of density at zero temperature
and different asymmetries.}\label{fig3}
\end{figure}

Now we consider asymmetric nuclear matter. The results of
numerical calculations for the energy gap as a function of density
for different asymmetries at zero temperature are shown in
Fig.~\ref{fig3}.
As one can see, with increasing  asymmetry the value of the energy
gap decreases and the density interval, where $np$ pairs exist,
shrinks in the direction of zero density. In reality solutions
exist for any $\al<1$ (the phase curves for larger values of $\al$
are not shown in Fig.~\ref{fig3}) and  corresponding density
interval contracts more and more to the point $\vrr=0$, when
asymmetry increases. To understand why the isospin asymmetry loses
its efficiency in destroying  $np$ pairing correlations in low
density region, let us note that at zero temperature the
contribution to the integral in the gap equation gives the
interval $[0,\ve_c]$, excluding the domain
$[\mu_{00}-\De\ve,\mu_{00}+\De\ve]$, where
$\De\ve=\sqrt{\mu_{03}^2-\De^2}$. In the weak coupling regime
($\mu_{00}\gg\De$) with increasing  asymmetry the width 
of this domain also increases, that results in considerable
reduction of the energy gap magnitude, until  it completely
vanishes. However, when the chemical potential passes through zero
 and becomes negative, only part of the window participates in suppressing the energy
gap, with the right end of the blocking interval going to zero at
$\vrr\rightarrow0$.

\begin{figure}[tbp]
\begin{flushleft}\includegraphics[height=7.2cm,width=14.4cm,trim=26mm 152mm
39mm 69mm, draft=false,clip]{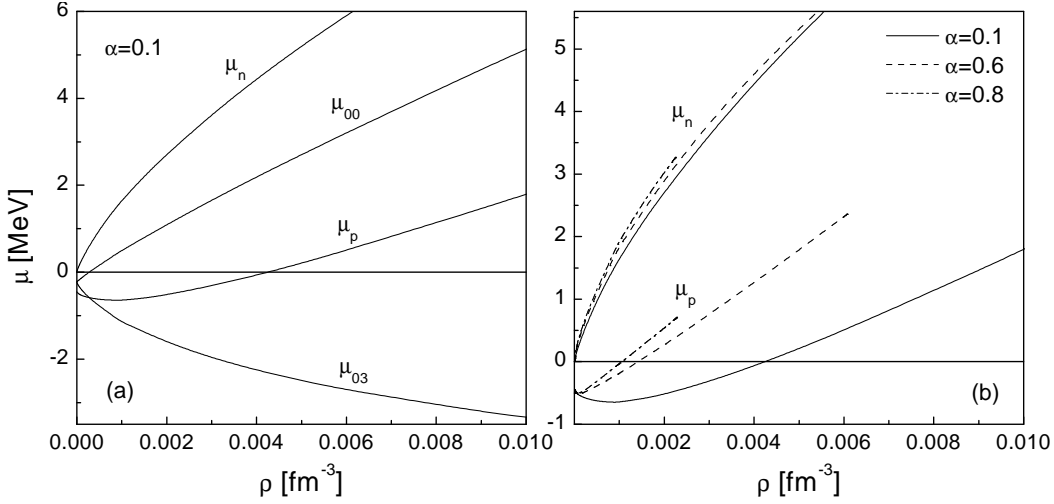}
\end{flushleft}
\caption{Various chemical potentials  as  functions of density at
zero temperature and different asymmetries: (a)
$\mu_{00},\mu_{03},\mu_{p},\mu_{n}$; (b) $\mu_{p}$ (lower curves),
$\mu_{n}$ (upper curves).}\label{fig4}
\end{figure}

In Fig.~\ref{fig4}(a) it is shown zero temperature behavior of
chemical potentials $\mu_{00},\mu_{03},\mu_{p},\mu_{n}$ as
functions of density at very low densities of nuclear matter and
finite isospin asymmetry. One can see, that at some critical
density the chemical potential $\mu_{00}$ changes its sign and
$np$ Cooper pairs smoothly go over into deuteron bound states. The
asymptotic behavior of chemical potentials at $\vrr\rightarrow0$
is $\mu_{00},\mu_{03}\rightarrow-\ve_b/2$, and, hence,
$\mu_{p}\rightarrow-\ve_b$, $\mu_{n}\rightarrow0$. This asymptotic
behavior does not depend on isospin asymmetry, that is confirmed
by the results of numerical calculations, shown in
Fig.~\ref{fig4}(b), where the density dependence of chemical
potentials $\mu_p,\mu_n$ is depicted for different values of the
asymmetry parameter.

\end{document}